# Design and Management of Vehicle Sharing Systems: A Survey of Algorithmic Approaches

Damianos Gavalas[1,4], Charalampos Konstantopoulos[2,4] and Grammati Pantziou[3,4]

[1] Department of Cultural Technology and Communication, University of the Aegean,
Mytilene, Greece
Email: dgavalas@aegean.gr

[2] Department of Informatics, University of Piraeus,
Piraeus, Greece
Email: konstant@unipi.gr

[3] Department of Informatics, Technological Educational Institution of Athens,
Athens, Greece
Email: pantziou@teiath.gr

[4] Computer Technology Institute & Press "Diofantus", Patras, Greece

**Abstract**

Vehicle (bike or car) sharing represents an emerging transportation scheme which may comprise an important link in the green mobility chain of smart city environments. This chapter offers a comprehensive review of algorithmic approaches for the design and management of vehicle sharing systems. Our focus is on one-way vehicle sharing systems (wherein customers are allowed to pick-up a vehicle at any location and return it to any other station) which best suits typical urban journey requirements. Along this line, we present methods dealing with the so-called asymmetric demand-offer problem (i.e. the unbalanced offer and demand of vehicles) typically experienced in one-way sharing systems which severely affects their economic viability as it implies that considerable human (and financial) resources should be engaged in relocating vehicles to satisfy customer demand. The chapter covers all planning aspects that affect the effectiveness and viability of vehicle sharing systems: the actual system design (e.g. number and location of vehicle station facilities, vehicle fleet size, vehicles distribution among stations); customer incentivisation schemes to motivate customer-based distribution of bicycles/cars (such schemes offer meaningful incentives to users so as to leave their vehicle to a station different to that originally intended and satisfy future user demand); cost-effective solutions to schedule operator-based repositioning of bicycles/cars (by employees explicitly enrolled in vehicle relocation) based on the current and future (predicted) demand patterns (operator-based and customer-based relocation may be thought as complementary methods to achieve the intended distribution of vehicles among stations).

## 1. Introduction

Sustainable principles in urban mobility urge the consideration of emerging transportation schemes including vehicle sharing as well as the use of electro-mobility and the combination of vehicle transfers with greener modes of transport, including walking, cycling and public transportation.

Bike-sharing programs have received increasing attention in recent years aiming at improving the first/last mile connection to other modes of transit and lessen the environmental impact of transport [1]. Bike-sharing programs are networks of public use bicycles distributed around a city for use at low cost. The programs comprise short-term urban bicycle-rental schemes that enable bicycles to be picked up at any bicycle station and returned to any other bicycle station, which makes bicycle-sharing ideal for point-to-point trips. The principle of bicycle sharing is simple: individuals use bicycles on an "as-needed" basis without the costs and responsibilities of bicycle ownership [2]. The earliest well-known community bicycle programme is launched in 1965 in Amsterdam, the Netherlands.

Current bike-sharing systems deploy bikes picked up and returned at specific locations (docking stations) and typically employ some sort of customer authentication/tracking (through the use of an electronic subscriber card) to avoid theft incidents [3]. Recent developments pave the way for next-generation bike sharing known as the "demand-responsive multimodal system" [2]. Such systems will emphasize on flexible docking stations (relocated according to usage patterns and user demands), incentivize user-based redistribution (by using demand-based pricing wherein users receive a price reduction or credit for delivering bicycles at empty dockings), enable integration of bike-sharing with public transportation and car-sharing locations (via smartcards, which support numerous transportation modes on a single card) and GPS-based tracking. Online services like Social Bicycles (SoBi) [4] allows users to locate, reserve, and unlock a bike with a smartphone app, while also employing a rewarding scheme to motivate cyclists to return bikes to central stations/hubs.

Similarly to bike sharing, car sharing is a model of short-term car rental, particularly attractive to customers who make only occasional use of a vehicle, enabling the benefits of private cars without the costs and responsibilities of ownership [5]. Car sharing first appeared in North America around 1994. Replacing private automobiles with shared ones directly reduces demand for parking spaces and decreases traffic congestion at peak times, thereby supporting the vision of sustainable transportation. Car sharing operators typically allow cars to be picked up from designated stations (depots) with customers required to return vehicles to their original pick up locations (such schemes are referred to as two-way car sharing systems). Most operators have been reluctant in introducing innovative features (e.g., one-way rentals, ridesharing) due to added management complexities [6].

These complexities were responsible for the failure of Honda's Diracc system in Singapore, one of the best-known one-way car sharing experiments in the world, after *6* years of operation (the system has been discontinued in 2008). Diracc failed mainly because it proved unable to maintain the quality of service (i.e. car availability) required by customers due to one-way trips leaving the system with significant imbalance in vehicle stocks. Indeed, during a typical day, the number of cars throughout a network shifts toward certain destinations; for instance, drivers commuting from the suburbs to downtown offices generate surplus of cars at certain stations, while depleting fleets at other stations. Nevertheless, some recent car sharing initiatives -notably, Daimler's Car2Go[1] and BMW's DriveNow[2] -offer the option of one-way car-sharing, as long as the customer drops off the car at any available public parking space within a designated operating area.

The design and management of a car sharing system raise several optimization problems. First, optimal fleet sizes along with the location of the parking stations should be determined [7]. Further on that, operators allowing one-way rides need to develop strategies to reallocate the vehicles and restore an optimal fleet distribution among stations. Such a distribution could respond to the short-term needs at a particular station or be based on an historical prediction (i.e. estimating future demand to proactively schedule relocations) [73]. While bike sharing operators typically employ dedicated vehicles for relocating bunches of bicycles to depots with depleted stock, vehicle relocation in car sharing programs is more demanding. In particular, the activities of vehicle relocation can be carried out by the user itself or by the operator. In the first case, the user is incentivized to car pool or to choose another location or reservation time; in the second case, which is currently more common, the vehicles are physically transported using dedicated trucks or personnel.

A recent development in vehicle sharing systems has been the employment of fully electric vehicles (EVs) as a means of lowering the environmental footprint of urban mobility. Further complicating things, the design of EV-sharing systems needs to consider two additional constraints: the availability of charging facilities on parking stations and the design of relocation strategies which take into account vehicles residual energy [8].

The above detailed challenges call for intelligent algorithmic solutions to support the long-term viability of vehicle sharing systems. Such algorithmic approaches should aim at the highest possible quality of service for customers and reduced capital investment for operators with respect to both system deployment and operating expenditures. To achieve these objectives, the whole range of deployment and operational parameters inherent in vehicle sharing systems should be carefully addressed: long-term strategic planning of systems, tactical decisions to enable user-based regulation to the benefit of the systems and operational issues.

This chapter offers insights on research tackling the above main issues related to the design and operation of public bicycle/car sharing systems. The focus is on mathematical models and algorithmic solutions developed so far, especially those that address cost and pricing models, depot location optimization, mobility and demand modeling, ways of balancing vehicle stocks across stations (i.e. relocation strategies) in one-way vehicle sharing systems. The objective is to identify the state-of-the-art along with possible paths for future developments in this field.

---

[1] https://www.car2go.com/
[2] https://de.drive-now.com/en/

The remainder of the chapter is structured as follows. Section 2 elaborates further on the challenges and objectives relevant to the design of vehicle sharing systems. Section 3 overviews models and algorithmic approaches for the design, operation and management of vehicle sharing systems. Section 4 presents algorithmic approaches for ride sharing. Finally, Section 5 provides insights on open issues and research challenges in the field while Section 6 concludes the paper.

## 2. Challenges and objectives in the design of vehicle sharing systems

Recent research analyzed the factors affecting the success of bike-sharing programs [9], [10]. These factors range from the built environment (infrastructures, facilities at work, etc.) to factors related to the natural environment (topography, seasons and climate or weather), socio-economic and psychological factors (attitudes and social norms, ecological beliefs, habits, etc.), and other factors related to utility theory (cost, travel time, effort and safety). Factors gaining growing interest involve bike station location, cycling network infrastructure (bike paths) and the operation of bicycle redistribution system [11]. The stations must be located in close proximity to one another and to major transit hubs and be placed in both residential (origin) and commercial or manufacturing (destination) neighborhoods, which makes bike-shares ideal as a commuter transportation system [12], [1]. Existing examples show that the bike stations should not be located more than 300-500m from important traffic origins and destinations. Given the complexity of bicycle facility planning and the importance of station distribution for operating bike-sharing programs, formal approaches are needed to model the problem variables and derive optimal solutions with respect to minimizing investment cost and maximizing utility for the users. Among others, optimal solutions should determine the number, location and capacity (in bikes and docks) of the stations as well as the bicycle lanes needed to be setup.

On the other hand, equally important for bike-sharing systems success is to guarantee bicycle availability. Each rental station must carry enough bicycles to increase the possibility that each user can find a bicycle when needed. Therefore, measures of service quality in the system include both the availability rate (i.e., the proportion of pick-up requests at a bike station that are met by the bicycle stock on hand) and the coverage level (the fraction of total demand at both origins and destinations that is within some specified time or distance from the nearest rental station). Due to the one-way rental policy, bikes are likely to get

- 4 -

stuck in areas of lower individual mobility demand (cold spots) while needed in zones of higher demand (hot spots). To make the system more efficient and more profitable, this imbalance of supply and demand could be adjusted by applying different intervention (i.e. relocation) strategies [13].

The need to ensure vehicle availability in high-demand areas is also acknowledged for car-sharing systems [14]. However, relocation of cars is more troublesome than that of bicycles (up to 60 bicycles can be transported altogether to hot spots on a bicycle carrier, contributing to cost and effort savings [15]). Some studies suggest the use of road vehicles (car carriers) with fully automated driving capabilities (typically moving along dedicated tracks), coordinated by centralized management systems, able to autonomously relocate to satisfy user demands [16]. Redistribution of vehicles may also be provided by a fleet of limited capacity tow-trucks located at various network depots; using such an approach the problem can be conveniently modeled as pickup and delivery problem [8]. However, dedicated transport trucks are of little use in most urban settings due to stations not easily reachable by heavy-duty trucks and the time consuming vehicle loading/unloading operations [17]. Thus, the scheme most commonly encountered in practice engages teams of employed drivers who undertake the relocation of vehicles thereby significantly increasing operational cost.

Recently, the decreased manufacturing cost of EVs along with their eco-friendly characteristics (fuel economy and lowered greenhouse gas emissions) has attracted the attention of car-sharing companies[3]. So far, the main body of EVs-relevant algorithmic research focuses on novel energy-efficient routing algorithms motivated by the unique characteristics of EVs (limited cruising range, long recharge times and the ability to recuperate energy during deceleration or when going downhill) [18], [19].

EV-sharing systems are also unique with respect to their design and operational requirements. Specifically,

(1) Sufficient battery availability at pick-up time should be ensured so as to travel reliably to user's destination [20].
(2) Vehicle relocation policies should take into account the energy availability of vehicles at stations, in addition to physical availability [21].

---

[3] Among other operators, Car2Go has launched (as of November 2011) an EV car-sharing network currently covering San Diego and Amsterdam. Through a user-friendly web interface, users interested in driving a shared EV, Car2Go members are be able to view the exact location of available EV along with their batteries state of charge and proceed to online reservations. If the battery performance sinks below 20%, the driver must end his/her trip at a charging station (found through an in-built navigation system

(3) Pick-up/drop-off locations are determined by the existence of charging stations (for instance, the 300 Car2Go vehicles and other EVs in Amsterdam have access to 320 charging stations in the city area).

(4) The anticipated transformation of urban parking stations to charge-park stations in support to EV power demands is expected to create considerable load on the power grid, hence, intelligent approaches are in need to flatten the load peak, thereby deferring investments in grid enhancement [22].

**3. Models and algorithmic approaches for optimizing vehicle sharing systems**

Bicycle and car sharing systems are complex dynamical systems with stochastic demand whose modeling and performance analysis is very important for their implementation and performance as well as for ensuring an effective regulation of vehicle traffic flows. Different approaches and methodologies have been proposed in the literature for modeling and studying design, operational and management issues of bicycle/car sharing systems. Such approaches include mixed integer programming approaches (e.g. [30], [28]), stochastic programming approaches ([29]), simulation methodologies (e.g. [73], [74], [75]). Although Petri nets have been a tool used rather successfully in the literature for modeling and evaluating the performance of dynamic and complex systems in various domains (e.g., traffic control of urban transportation systems [39], [40], [41], [42] and planning [43], [44], [45]), very limited research work exists in Petri net models for modeling and performance analysis of bicycle and car sharing systems ([25] and [46]).

Besides OR approaches (using either mathematical programming or Petri nets and closed queuing networks) that support decision making in the design and management of bicycle and car sharing systems, data mining techniques have also received attention in the literature. Data mining is particularly suitable to analyze and predict the dynamics of such systems. The analysis of the temporal human mobility data in an urban area (using the amount of available bicycles/cars residing in the stations of vehicle sharing systems) may offer insights on the system structure and operation; therefore, statistical and prediction models can be developed for the tactical and operational management of these systems. Some of the research works focusing on using data mining to analyze bicycle sharing systems are the following:

- In [47] Froehlich et al. provide a spatio-temporal analysis of data collected for the number of available bikes and vacant bike stands from Barcelona's bicycle sharing system. Stations are clustered according to the number of available bikes and an activity score assigned in the course of day. Then, visualization is used to identify shared behaviors across stations and show how these behaviors relate to location, neighborhood and time of day. The authors show that fairly simple predictive models are able to predict station usage with an average error of only two bicycles and can classify station state (i.e., full, empty, or in-between) with 80% accuracy up to two hours into the future.

- In [24] Kaltenbrunner et al. detect bike usage patterns in data from Barcelona's bicycle sharing system. Their results are similar to those of Froehlich et al [47]. They present a statistical model that predicts the number of free bikes and vacant bike stands at stations some minutes ahead in time.
- Borgnat et al. [48] use data mining to analyze the dynamics of bike movements in Lyon's bike sharing system. Temporal patterns in the system-wide bike usage are examined. Weekdays show usage peaks in the morning, at noon and late afternoon, whereas usage is concentrated in the afternoon on weekend days. A statistical model for the prediction of the number of rentals on a daily and hourly basis is proposed. Furthermore, spatial patterns are examined by clustering bike flows between stations. Spatial and temporal dependencies exist between stations of clusters interchanging many bicycles.

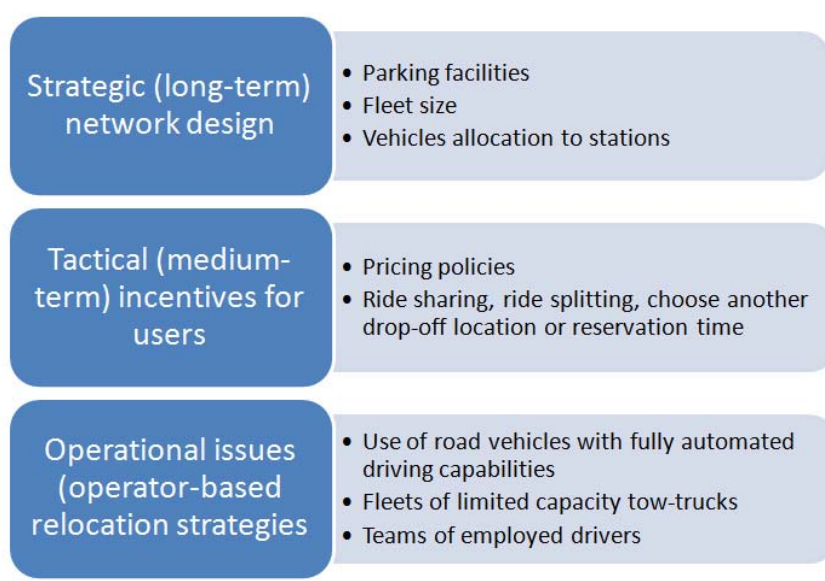

**Figure 1:** Main issues related to the design, management and operation of vehicle sharing systems.

Vogel et al. [23] identify three main issues related to the design, management and operation of bicycle/car sharing systems. The proposed design and management measures (aiming at alleviating imbalances in the availability of bicycles/cars) are distinguished into three separate planning horizons (see Figure 1):

(1) Strategic (long-term) network design comprising decisions about the location and the number of stations as well as the vehicle stock at each station.

(2) Tactical (mid-term) incentives for customer-based distribution of bicycles/cars i.e., incentives given to users so as to leave their vehicle to a station different to that originally intended (this may be regulated through pricing schemes adaptable to the system state). For example, Figure 2 illustrates two possible options offered to the user willing to move from a location *A* to a location *B*: The $A \rightarrow S_1 \rightarrow S_3 \rightarrow B$ option (i.e., the user leaves the vehicle at the station $S_3$ and walks to the destination location *B*) is the shortest time one, while the $A \rightarrow S_1 \rightarrow S_2 \rightarrow B$ option (i.e., the user leaves the vehicle at the station $S_2$ and walks to the destination location *B*) is associated with the incentivized scheme.

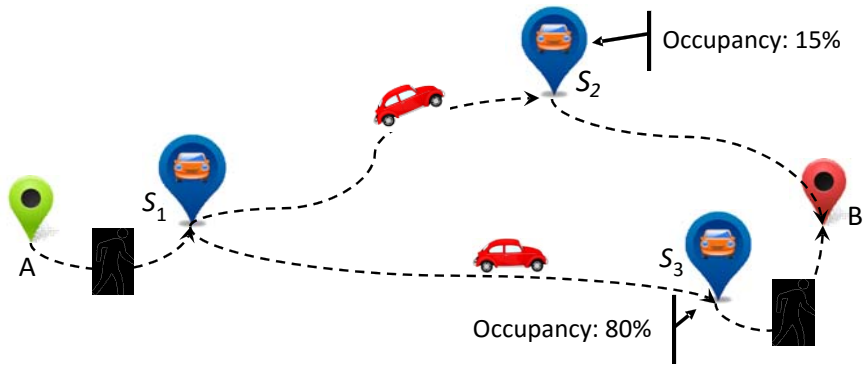

**Figure 2**. Illustration of the different options offered to a user moving from *A* to *B*

(3) Operational (short-term operator-based) repositioning of bicycles/cars based on the current state of the stations as well as aggregate statistics of the stations' usage patterns. For example, Figure 3 illustrates a relocation plan for a vehicle sharing system, based on the system data shown in Table 1.

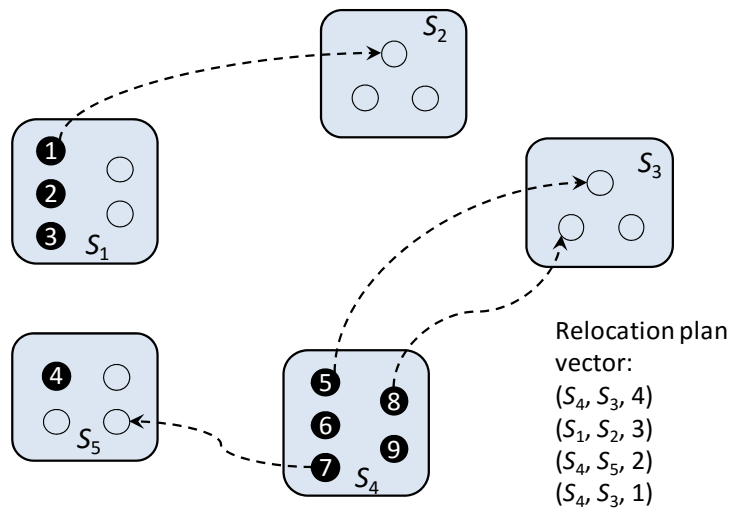

**Figure 3**. Illustration of a relocation plan. Circles filled with black color represent parked cars while empty circles represent empty parking spaces.

| Stations | $S_1$ | $S_2$ | $S_3$ | $S_4$ | $S_5$ |
|---|---|---|---|---|---|
| Capacity | 5 | 3+ | 3 | 5 | 4 |
| Current occupancy | 3 | 0 | 0 | 5 | 1 |
| Targeted occupancy | 2 | 1 | 2 | 2 | 2 |
| Surplus/ Deficit | +1 | -1 | -2 | +3 | -1 |

**Table 1**. Example snapshot showing the capacity, current/targeted occupancy and surplus/deficit in vehicles.

In the sequel of this Section, we overview algorithmic solutions proposed for the strategic design of vehicle sharing systems (Subsection 3.1), present modeling approaches supporting pricing schemes and incentives for customer-based distribution of vehicles (Subsection 3.2) and summarize algorithmic approaches for the problem of operational repositioning of vehicles (Subsection 3.3).

### 3.1 Algorithmic approaches on the strategic design of vehicle sharing systems

***Integer programming based approaches.*** Lin and Yang [26] have been the first to investigate the problem of strategic design of bicycle sharing systems. The problem investigated is the following: given a set of origins, destinations, candidate sites of bike stations and the stochastic travel demands from origin to destination, the problem's output comprises the location of bike stations, the bicycle lanes needed to be setup and the paths to be used by users from each origin to each destination, the objective being to minimize the overall system cost. The authors take an integrated view of the system cost, considering both the user's and the investor's point of view. In particular, the investor's cost comprises the facility costs of bike stations, the setup costs of bicycle lanes, bicycle stock and safety stock (for serving the demand at peak hours) costs. The level of service provided to the user is measured by the demand coverage level (defining penalty costs for uncovered demands) and travel costs (for both walking and cycling). The problem has been formulated as an integer nonlinear program.

Martinez et al. [27] formulated a mixed integer linear program (MILP) aiming to optimize the location of shared biking stations and the fleet dimension. This study also considered bike relocation operations among docking stations (the relocation operations cost is considered as an additional system cost factor, yet not explicitly included as a decision variable in the MILP formulation). A general model framework has been proposed, which computes several days of operation, maintaining the dimensioning data from previous iterations, re-computing the hour operation MILP model and updating the system design, until the configuration reaches a net revenue equilibrium, producing a stable and optimal system configuration.

Correia and Antunes [28] addressed the optimization problem of selecting sites for locating depots in order to maximize the profits of a one-way car sharing organization. Revenues are generated from renting the vehicles against some price rate while several types of expenses are considered (maintenance costs for vehicles and depots, vehicle depreciation costs and vehicle relocation costs). Relocation operations are only considered at the end of the day, unlike previous studies wherein the main emphasis was on optimizing such operations [29, 30]. Three mixed integer programs (MIP) have been modeled which determine the optimal number, location, and capacity for the depots, each one corresponding to a different trip selection policy. According to the first policy, the operator is free to accept or reject trips in the period they are requested according to the profit-maximization objective; the second policy assumes that all trips requested by clients are approved; the third policy allows a trip request to be rejected in the case that there are no vehicles available at the pick-up depot. The optimization models have been tested in a case study involving the municipality of Lisbon, Portugal.

Boyaci et al. [31] proposed a generic model for supporting the strategic (number and location of required stations) and tactical (optimum fleet size) decisions of one-way car-sharing systems by taking into account operational decisions (i.e. relocation of vehicles). The authors formulated a mathematical model (integer program) and conducted sensitivity analysis for different parameters. The objective function seeks to maximize the overall profit which considers the revenue generated from vehicle rentals in addition to user costs (proportional to the time required to reach the origin station from the start location and the end location from the destination station) and system costs (unserved customer cost, vehicle operating cost, station opening cost and relocation cost). The proposed model has been applied for planning and operating a station-based EV-sharing system in the city of Nice, France.

***Heuristics.*** Recognizing the complexity of the bicycle sharing system design optimization which precludes exact solutions for instances of realistic size, Lin et al. [32] approached the system's design as a hub location inventory problem[4] [33] that takes the coverage level into consideration and proposed a greedy algorithm for solving it efficiently. The greedy-drop heuristic iterates between locating bicycle stations given a collection of bicycle lanes, and locating bicycle lanes given a set of bicycle stations. In particular, all candidate stations and the bicycle lanes connecting them are initially marked as "open". The algorithm then iteratively removes the currently open station, which if closed, would result in the largest total cost reduction, until no further cost reduction is possible. Likewise, bicycle lanes are removed, as long as their removal results in cost reduction. The overall solutions cost is calculated utilizing the mathematical cost model introduced in an earlier study [26]. When testing the algorithm in test instances for which enumeration is possible, the heuristic solution has been found within a 2% gap from the optimal.

---

[4] The hub location problem has been one of the important classic facility location problems. Hub facilities concentrate flows to achieve economies of scale. Flows between origins/ destinations and hubs and between pairs of hubs are consolidated into a smaller set of links rather than serving demand with direct links. The hub location problem involves determining the hub facilities and determining the links to connect origins, destinations and hubs.

Kumar and Bierlaire [34] developed an optimization model to identify the most appropriate locations for establishing car-sharing stations such that the overall system performance is maximized (the main measure of stations performance is the average number of rides per day). The model considers car-sharing systems exclusively allowing round-trips. The model balances between the estimated attractiveness of key demand drivers in a locality and the locality's proximity to an existing station. The authors first build a linear regression model (applied on historical data of the Auto Bleue EV-sharing operator, in Nice, France) to identify the key demand factors that affect stations performance[5]. Then they formulate a mixed-integer quadratic (higher-order) program. The objective of the mathematical model is to maximize the combined performance of all selected stations ($n$ stations are selected among $k$ candidates; candidate stations are assumed to be located at the centroid of pre-specified city localities). The main trade-off decision made by the model involves locating more stations in "highly attractive" localities versus locating new stations in "less attractive" but untapped localities (establishing too many stations at attractive locations does not increase the overall system performance as they tend to cannibalize each other's performance). Last, a heuristic is proposed to solve the problem. The heuristic first estimates the "best performing stations" based on all parameters except distance and public transport ridership. In the first iteration, all $k$ stations are assumed to be operational; the contribution of public transportation and distance is then computed. In the next step, the $n$ best locations are picked to place the stations. Now the public transportation and distance contribution is recomputed assuming that only these n proposed stations are operational. Based on the changes in the objective function, the $n$ best locations to place the stations are again selected. This process is repeated until the selected set of n stations remains unchanged.

The problem of determining the fleet dimension (size) and the distribution of vehicles among the stations of a car-sharing system was studied in relation to electrically powered one-person vehicles (Personal Intelligent City Accessible Vehicles, PICAVs), which enable accessibility for all in urban pedestrian zones [35]. This system allows one-way trips among stations (parking lots that offer vehicle recharging services) located at inter-modal transfer points and near major attraction sites within the pedestrian area. The number, the location, and the capacity of the stations are not determined by the model. To cope with the imbalanced accumulation of the one-way system, this model enrolls a human supervisor. The task of the supervisor is to direct users that are flexible in returning the car to alternative stations, as to achieve a balanced operation and fulfil a maximum waiting time constraint. The cost minimization problem has been solved using a simulated annealing-based approach (the cost function takes into account both the transport system management and the customer cost, i.e. the cost of vehicles and the total customer waiting time, respectively).

---

[5] Stations performance have been found to increase with the share of high income/education population (in the locality), the share of public transport ridership, the share of car usage to reach workplace, the presence of mobility attractors (mainly commercial centers, hotels and colleges), the population density, the presence of transit hubs; on the contrary, distance from customers residence decreases the performance of stations.

***GIS-based approaches.*** Geographic Information Systems (GIS) represent a highly useful tool for determining bike station locations. Larsen et al. [36] presented a GIS-based grid-cell model to identify and prioritize cycling infrastructure investments using the example of Montreal, Canada. The main result is a grid-cell layer of the study region wherein high-priority grid-cells represent those areas most appropriate for bicycle infrastructure interventions (i.e. the areas where new cycling facilities would provide the maximum benefit to both existing and potential cyclists). Rybarczyk and Wu [37] used GIS-based multi-criteria decision analysis both to evaluate the quality of bicycle facilities utilizing supply and demand-based objectives. Analyses were conducted at two levels: network (bicycle facility) level and neighborhood level. Network level analysis can address site specific issues and provide detailed information for further improvements. By contrast, neighborhood level analysis provides a strategic view of bicycle facilities in an urban area, and facilitates policy development and implementations.

Garcia-Palomares et al. [38] proposed a GIS-based method to calculate the spatial distribution of the potential demand for trips, locate stations using location-allocation models, determine station capacity and define the characteristics of the demand for stations. The authors follow a four-step approach: First the distribution of the potential user demand is assessed (the number of trips generated and attracted for each transport zone is calculated based on the population and employment associated with each building). The location-allocation models ($p$-median and maximum coverage[6]) are then applied defining obligatory bike-stations, candidate locations, the number of stations to be located and the type of solution chosen. Once bike-station locations and potential demand upon stations are obtained, the stations capacity (number of bicycles and docks) is calculated; also, the stations are characterized (as trip generators or attractors) making it possible to vary the number of bicycles according to the time of day, leading to more efficient bicycle redistribution systems. The final step is the analysis of stations in terms of accessibility (a measure of usefulness, which considers the volume of demand allocated to the station and its distance to the potential origin/destination stations of the users); this way, it is possible to prioritize stations within the bike-sharing program (eliminating those with poor accessibility).

---

[6] In the maximum coverage location-allocation model, the stations are located such that as many demand points as possible are allocated to solution facilities within the impedance cutoff (200m).

***Data mining - based approach.*** Vogel et al. [23] use geographical information technology and data mining methods to gain insight into bicycle stations operations and try to incorporate this knowledge in the design of bicycle sharing systems (strategic and operational planning). In a case study, collected data related to the activity of the bike stations are provided as input to a data mining phase where cluster analysis is used to group stations according to their pickup and return activity patterns. The analysis reveals spatio-temporal dependencies of pickup and return activities at stations which support the hypothesis of Vogel et al. that usage patterns at bike stations and the type of customers using certain stations depend on the stations' location. Note that if the hypothesis holds, then usage patterns for already existing stations can be mapped to potential stations based on their locations. Therefore, strategic decisions about the bicycle sharing system can be supported.

The previous algorithmic approaches are summarized in Table 2 where a classification is also given according to whether they concern bicycle or car sharing systems.

|  | **Bike sharing systems** | **Car sharing systems** |
|---|---|---|
| Integer programming based approaches | Lin and Yang [26] <br> Martinez et al. [27] | Correia and Antunes [28] <br> Boyaci et al. [31] |
| Heuristic approaches | Lin et al. [32] | Kumar and Bierlaire [34] <br> Cepolina and Farina [35] |
| GIS-based approaches | Larsen et al. [36] <br> Rybarczyk and Wu [37] <br> Garcia-Palomares et al. [38] |  |
| Data mining based approach | Vogel et al. [23] |  |

**Table 2.** Algorithmic approaches on the strategic design of vehicle sharing systems

### 3.2. Tactical incentives for bicycles/cars distribution

In [49], a bike sharing system is modeled as a stochastic network and its steady state performance is analyzed using the mean field theory. Specifically, in this model, there are *N* bike stations, each of which can keep at most *K* bikes. Initially, there are *s* bikes in each station and therefore the total number of bikes in the system is *sN*. It is also assumed that the arrival rate at each station is *λ* (symmetric case) and the travel time between any two stations follows the exponential distribution with parameter *μ*. The authors determine the proportion of problematic stations (empty or saturated) at steady state. Specifically, they prove that the proportion of problematic stations at steady state is minimal when *s = K/2+ λ/μ* and the minimum is equal to *2/(K + 1)*. This is not an encouraging result, since for achieving low proportion of problematic stations, large capacities in the stations are needed, which is not always feasible due to space constraints and construction costs. The authors also show that the situation does not improve even when the users are aware of the problematic stations and they always pick one of the remaining stations for getting and leaving a bike. The performance of the system is greatly improved if simple incentives for the users are adopted. Specifically, the authors test the case when the user selects a station at random for leaving its bike and then s/he finally selects the least loaded. The analysis and the simulation results clearly show that the proportion of problematic stations is now much lower. This also holds in the case that only a fraction of users accept to follow that policy. Then, the authors study the asymmetric case where there are two clusters of stations and the customer arrival rate at the stations of one cluster is higher than at the second cluster. In this case, the performance of the system is much worse than in the symmetric case when there is no regulation mechanism for the bike distribution across the stations. Even the above incentive of two choices is not that effective in this case. So, the authors propose bike repositioning using a number of tracks. Indeed, the simulation results demonstrate much higher performance in the steady state if the trucks regularly redistribute the bikes across the stations.

In [50], a bike sharing system is presented where periodic redistribution of bikes across the stations is carried out by using a number of trucks and also incentives are given to the users to leave their bike to a different than the originally intended station. Incentives are regulated through a pricing scheme which is changing online according to the current state of the system. First, the authors use historic data for building user demand statistics of the bike sharing system. Specifically, they determine the average arrival and departure rate of customers at each station for a number of time intervals on each day differentiating between working days and weekends. Then, periodically, each time for a fixed planning horizon, they determine the truck routes for optimal redistribution of bikes across the stations. For the problem formulation, the authors assume deterministic flows in the network and also define a utility function at each station which determines the benefit of removing or adding bikes at the station at the current time with respect to the increase of the percentage of users whose requests will be satisfied at this station in the near future. Then, they study the problem of finding the best route for the case when only one truck is used. They also assume that during each trip, the single truck can visit at most a small constant number of stations. Then, they use a greedy approach and they build a tree emanating always from a specific spot (named maintenance depot in the paper); a number of stations are added iteratively to the tree so that the increase in the utility function over the additional cost incurred for reaching the station is relatively high. Having constructed the tree, a separate optimization problem is solved for each different route starting from the root of the tree and ending at the tree leaves. This optimization problem which is in the form of a quadratic program, refines the truck loading actions across each route leading to a more effective solution. Then, the authors generalize their solution for the case of multiple trucks in the system. Essentially, they follow a sequential approach fixing the route of trucks one after another. Finally, the authors study the problem of determining the pricing scheme which will have the lowest monetary cost while keeping the bike distribution across the stations at an optimal level. The basic assumption in their approach is that the users are rational thinkers and when the system proposes to them an alternative nearby station to leave their bikes, the users weigh the monetary reward they are going to receive for this choice against the monetary cost of travelling additional distance. For determining the best pricing policy, the problem is formulated as a problem of Model Predictive Control. More precisely, the best prices are determined for each different time step within a finite time horizon and then only the prices concerning the current time step are finally adopted. At the next time step, the problem is resolved since the system state may have changed in the meantime.

In [7], a vehicle sharing system is modeled as closed queuing network. The authors make the simplifying assumptions that the users always find parking space at the destination station and also when they do not find a vehicle at the origin station, they simply leave the system. By regarding the vehicles as the customers of the closed queuing network system, each parking station is viewed as a single server node with FIFO service policy and the service time is equal to the inter-arrival time of users at that station. The user arrival at each station is modeled as a Poisson process. It is also assumed that the network of parking stations is complete and thus there is a direct link for each pair of parking stations. Each vehicle at the origin station leaves that station along a certain outgoing link with a specific probability. The travelling time between two stations *i* and *j* is exponentially distributed with parameter $1/\mu_{ij}$. Each link *(i,j)* of the station network is modeled as a node with infinite number of servers and with total service rate equal to $n\mu_{ij}$ where *n* is the number of vehicles travelling along that link. The main objective in their analysis is to determine the optimal number of vehicles (fleet size) in the system such as the overall profit is maximized. In estimating this profit, the authors consider the revenue per unit time obtained from a vehicle rent by a user. They also take into account a maintenance cost per vehicle and an unavailability penalty when a customer cannot find vehicles available at a station. Then, they prove that the profit function is a concave function and its optimization derives two solutions at most. Next, they use mean-value analysis, for estimating these solutions.

In [51], the authors analyze a pricing scheme by modeling a vehicle sharing as a closed queuing network, basically following the approach in [7]. However, now, each station is assumed to have finite capacity and also the demand for each out-link of a station is elastic influenced by the price that should be paid for travelling along this link. It is also assumed that when a user picks a car at the origin station, the system ensures that there will be free parking space at the destination station by making reservation in advance. For determining the best pricing scheme, the time is partitioned into a number of time slots whose duration follows a certain distribution. In addition, the authors assume that the system has periodic behavior and the prices for each link should be determined only for the time slots within a single period of the system. Essentially, the whole problem is reduced to a Markov decision process wherein the set of actions applied at each moment should be determined. Apparently, this set of actions is the prices set for each link, which in turn affects the use of this link by the users of the system. Due to the huge state space of this process, the problem of obtaining the best pricing scheme cannot be solved in reasonable time. For this reason, the authors propose an approximation based on the fluid model where the stochastic demands are replaced by continuous flows with deterministic rate. Then, the problem is reduced to a continuous linear problem whose solution maximizes the sum of demands at each link of the station network.

In [52], the authors assume a vehicle sharing system where stations have unlimited capacity and the travel time between any two stations is negligible. These two assumptions simplify the modeling of the vehicle sharing system as a closed queuing network. Similarly to the previously discussed approaches, each station is a node of the closed queuing network where the jobs to be served are the cars at this station. The service rate of the server is equal to the rate of the customer arrival at that station which is modeled as a Poisson process. For each pair of stations there is a demand rate for the link connecting the corresponding nodes; this demand is leveraged by the price set for making a trip along this link. However, no method is proposed for adjusting these prices to maximize profit. Actually, the authors study the problem of finding the link demands which maximize the number of trips sold. Also, for each link, there is a separate upper bound for the demand passing through that link. This bound is implicitly determined by the lowest price that the system operator will set for the corresponding link. Then, the authors solve the maximum circulation problem on a flow network which results from the queuing network by viewing the upper bounds on the link demands as the edge capacities on this flow network. Note that in the maximum circulation problem, there is no source and sink node, and the objective is to maximize the circulated flow in the network without violating the capacity constraints. As the solution of this problem may yield zero flows for some links, the resulting flow network may be disconnected with a number of strongly connected components. Then, for each strongly connected component, the availability of each station at that component is determined, that is the probability that a new customer will find a vehicle at that station. Apparently, this probability is a function of the number of vehicles and the number of stations at the component. Given a specific distribution of vehicles across the different strongly connected components, the expected number of trips taking place in the system can now be calculated from the solution of the maximum circulation problem and the previously estimated station availabilities. Next, the authors give a greedy algorithm for determining the distribution of the vehicles across the strongly connected components mentioned above, which maximizes the expected number of trips sold. They also prove that this greedy approach is actually optimal based on the fact that the expected number of trips is a concave function of the number of vehicles within each strongly connected component. Finally, they present some preliminary results about the approximation ratio of their approach. Specifically, they claim, without a complete proof, that the proposed policy is a tight *N/(N+M−1)* approximation on both static and dynamic optimal policies where *N is* the total number of vehicles and *M* is that number of stations of the vehicle sharing system.

The deterministic version of the above problem is also studied in [53]. In this setting, the trips planned to take place in a fixed horizon are known in advance. Similarly to the above approach, each link is associated with a fixed price to pay for following that link. In addition, for each trip, users set a maximum price they are willing to pay. A trip is cancelled, if the price of this trip's link is higher than the maximum price for that trip. The optimization problem in this scenario is to determine the prices at each link so that the total system revenue is maximized. The authors prove that this problem as well as a number of variants are all NP-hard problems.

In [46] a user-based solution for the vehicle relocation problem in car sharing systems is proposed. In particular, an approach of using rental pricing incentives is presented and assessed. Incentives are intended to influence the travel behavior of the users according to the system conditions, monitored in real time. The proposed solution is based on a model of an electric-car sharing system developed in a Timed Petri Net (TPN) framework. Note that TPNs use graphical and mathematical descriptions to represent both the static and the dynamic aspects of the modeled system; the graphical representation enables a concise way to design and verify the model, while the mathematical description allows simulating the system in software environments, by considering different dynamic conditions [46]. The proposed vehicle relocation strategies have been applied to the real case of the electric-car sharing system of Pordenone, Italy. The simulation results show that a system which ignores the operative conditions of the service suggesting always to its customers to return the vehicles as soon as possible, does not lead to the rebalancing of the number of vehicles parked in each station. On the other hand, giving incentives to the users which depend on the real time monitoring of the system, can increase the number of served customers and, therefore, improve the overall system performance. The results also show that the effectiveness of the proposed solution decreases as the congestion level of the system grows highlighting the limits of such an approach.

In [76] the authors present a crowdsourcing mechanism that incentivizes the users in the bike repositioning process by providing them with alternate bike pick up or drop off locations in exchange for monetary incentives. The main component of the system is the Incentives Deployment Schema (IDS) that handles the user's request through a Smartphone App. The IDS communicates with the bike sharing system infrastructure to evaluate the current and predicted status of the stations, and decides whether to offer on not incentives to the user. In order to maximize the efficiency under given budget constraints, the authors design a dynamic pricing mechanism using the approach of regret minimization in online learning that can learn over time about the optimal pricing policies. The users are considered as strategic agents who may untruthfully report information about their personal cost and location to maximize their profit.  The pricing mechanism DBP-UCB (Dynamic Budgeted Procurement using Upper Confidence Bounds), is a dynamic variant of BP-UCB presented in [77]. The proposed system is evaluated through simulations using historical and user survey data. Finally, the system was deployed on a real-world bike sharing system for a period of 30 days in a city of Europe, in collaboration with a large scale bike sharing company. According to the authors this is the first dynamic incentives system for bikes repositioning ever deployed in a real-world bike sharing system.

The previous algorithmic approaches are summarized in Table 3 where a classification is also given according to whether they concern bicycle or car sharing systems.

|  | **Bike sharing systems** | **Car sharing systems** |
|---|---|---|
| Stochastic network modeling approach | Fricker and Gast [49] | |
| Model Predictive Control approach | Pfrommer et al. [50] | |
| Closed queuing network modeling approach | | George and Xia [7] <br><br> Waserhole and Jost [51], [52] <br><br> Waserhole et al. [53] <br><br> Briest and Raupach [66] |
| Timed Petri Net | | Clemente et al. [46] |

| modeling approach | | - | |
| Regret Minimization approach | Singla et al. [76] | | |

<div align="center">**Table 3**. Algorithmic approaches on tactical incentives for bicycles/cars distribution</div>

### 3.3. Operational repositioning of bicycles/cars

In a bike-sharing system, there is a set of stations providing bicycles for rent, each with a specified capacity of allowed bicycles. A customer may rent a bicycle at a station, use it for a period of time and then leave it to another station. Since, the stations have a specified capacity and the number of bicycles available for rent is restricted, shortage events may occur. A shortage event occurs when a customer tries to rent a bicycle from an empty station or tries to return a bicycle in a full station [61]. To eliminate shortage events, hence customers' dissatisfaction, it is necessary to reposition bicycles using a fleet of dedicated vehicles. The repositioning can either be static [61] i.e., it can take place during the night when no customer asks for bicycles or dynamic [62] i.e., occur during the day in order to remove bicycles from full stations and transfer them to stations with lack of bicycles. Two main factors are considered in a repositioning process, the number of vehicles removed/transferred to a station to meet the customers' need and the operational cost of the fleet of vehicles performing the repositioning. In several applications the latter factor may be considered insignificant compared to the impact of a dissatisfied user and hence it may be disregarded.

Chemla et al. [63] study the static rebalancing problem in bike sharing systems. The authors formulate the problem as a Single Vehicle One commodity Capacitated Pickup and Delivery Problem. In this formulation a single capacitated vehicle balances the stations transferring bikes from stations with excess of bikes to stations with shortage. The relocation is assumed to be static i.e., taking place during the night when there is no demand for bikes. The problem aims at producing a minimum cost vehicle route accompanied with loading/unloading number of bikes at each station. At route completion time all stations have to contain a predefined target number of bikes. The authors propose an intractable exact model for the problem. Then, the model is relaxed, obtaining an integer program with exponential number of constraints. This program is solved using a branch-and-cut algorithm, producing a lower bound for the solution. Apart from this approach, a tabu search heuristic is proposed to produce feasible solutions. The tabu search algorithm incorporates four different neighborhood structures. The tabu list contains a number of arcs previously removed during the execution of local search steps. Two different approaches are considered for the construction of the initial solution of the heuristic algorithm. In the first, a solution is constructed using a greedy heuristic procedure. In the second, an initial solution is obtained based on the solution of the integer program. The executed algorithms are the branch-and-cut algorithm for the integer program as well as the two versions (according to the construction of the initial solution) of the tabu search heuristic. The test instances used for evaluating the algorithms are based on the work in [64]. The experimental results indicated that tabu search incorporating the solution of the integer program produces higher quality solutions i.e., solutions with lower cost than the greedy initialization heuristic while the latter approach executes faster. Furthermore, the results indicate that the tabu search heuristic obtains, in general, solutions with cost close to optimal, achieving on average at most *5%* gap.

Raviv et al. [61] study the static repositioning problem of bicycles performed during the night using a fleet of vehicles. The problem aims at producing vehicle routes for bicycle repositioning in order to minimize a cost function. The cost function considered is a weighted combination of a convex penalty function based on the expected number of shortage events per station of the next day and the travel cost of the vehicle routes. Two Mixed Integer Linear Programming formulations of the problem are proposed, namely an arc-indexed formulation and a time-indexed formulation, each with different underlying assumptions. In the arc-indexed formulation a vehicle cannot visit a station twice, while no waiting is allowed at a station. These assumptions significantly reduce the number of decision variables and, hence, make the approach efficiently solvable. In the time-indexed formulation, the time period is discretized into small time periods and the decision variables taken into account extend the decision variables of the former formulation adding one more index, the time index. Furthermore, the restrictions of the arc-indexed formulation do not apply anymore. In this way the solution space of the latter formulation extends the solution space of the former. Since solving these programs would take a lot of computational time, a two-phase heuristic approach is considered. In the first phase the program relaxes the restriction of integer number of bicycles removed and transferred to stations, hence concentrating on the design of the vehicle routes. In the second phase, the program is solved with the integral restriction to the number of bicycles, with the decision variables concerning the vehicle routes treated as constants based on the solution obtained from the first phase. The algorithms have been tested on data from Paris (Velib system) consisting of at most 60 stations and two vehicles and Washington DC (Capital Bikeshare) consisting of 104 stations and two vehicles. The results indicated that the arc-indexed formulation combined with the two-phase heuristic approach was the most efficient approach yielding higher quality results in the allowed two hours execution time.

Weikl et al. [15] study the relocation problem of cars in free-floating car sharing systems. The relocation strategies are categorized as user-based and operator-based. In the former, the relocation is performed by the customers. Incentives and bonuses are offered to the users to either change their destination, leaving the rented car in a station with shortage of cars or share a car with other customers with similar trips. In the latter, the relocation is performed by the employees of the system, transferring cars from stations with excess to stations with shortage. The first approach is very profitable for the system, since no cost for car transferring is added, however customers may refuse to changer their trip or share a car. The second approach adds cost to the system, requiring employees' actions and car movement. Nevertheless, it is more reliable. Then a user-based algorithm of Di Febbraro et al. [65] and an operator-based algorithm of Kek et al. [30] are described to illustrate the different approaches. Finally, a two-step algorithm for car relocation in car sharing systems is introduced. In the first, offline step, a set of demand scenarios is produced based on real collected data. For each scenario, the optimum number of cars per station is computed and a set of relocation strategies is produced. In the second, online step, the number of vehicles currently placed in stations is compared to the optimum, computed in the current demand scenario. If these quantities differ, the appropriate relocation strategy produced in the previous step is applied.

The modeling of a car-sharing system as a closed queuing network is followed in [66], similarly to the works surveyed in Section 3.2. Again, the cars are considered as the pending jobs of the system and each parking station is viewed as a single server with the available cars at the station waiting in a queue for the next customer to come. Once more, the service rate of a server is essentially the inter-arrival rate of the customer arrival Poisson process at that server/station. The authors also assume that a customer picking a car at a station may drive to any other station with a certain probability. In addition, a redistribution policy is implemented wherein the staff of car-sharing company relocates cars so as to achieve maximum total profit. Specifically, a reward is credited when a customer uses a car for travelling between two stations, with the reward being proportional to the travelled distance. Similarly, a cost applies when a car is relocated by the company staff for achieving balanced car distribution across the stations. Again, this cost is proportional to the distance travelled for this relocation. It is also assumed that the reward value is higher than the relocation cost for the same travelled distance. Now, the overall objective is to determine the relocation policy which maximizes the total profit of the system. The problem is formulated as a linear programming problem and the optimal solution determines the average number of cars moving between each pair of stations due to customer requests and due to relocation which yields the highest net profit. Based on the optimal solution of this problem, a relocation policy is then determined. Namely, after a car arrives at a new station after completing a customer trip, the car is immediately relocated to a random target station according to a certain probability distribution. Specifically, the probability $p_{uv}$ of relocating a car from a station $u$ to a station $v$ is equal to $m_{uv}/y_u$ where $m_{uv}$ is the average number of cars relocated from $u$ and $v$ and and $y_v$ is the average number of cars at node $v$ after a customer request has been served at that station. The values of these two parameters derive from the optimal solution of the linear program discussed above. Now, the authors prove that this relocation policy yields profit within a factor of 2 of the optimal policy's profit. They also prove, via a reduction from the Set Packing Problem, that finding the optimal relocation policy in the car sharing problem is an APX-hard problem, in general. Finally, they provide some preliminary results for the discrete-time version of the car sharing problem where customers are not arriving according to a Poisson process but simultaneously at all nodes at regular intervals. Also, after each round of customer requests, a relocation policy may relocate all cars regardless of whether they were moved due to a customer request. Then, the authors study the problem assuming that the distance between any two stations is 1 and that a customer at a station will select the destination station uniformly. In this case, the optimal policy is proved to be not performing any car relocation. Then, the average fraction of non-empty queues in the system is determined and this is also the approximation ratio with regards to the optimal policy.

Gendreau et al. [67] tackle the problem of dynamically relocating emergency vehicles in order to cover the most possible population. For example, when a vehicle leaves its location for a service, the remaining vehicles are relocated to be able to cover as much population as possible. The problem is formulated as a Maximal Expected Coverage Relocation Problem (MECRP). The input of the MECRP consists of *n* vehicles, a directed edge-weighted graph *G = (V, A)*, with *V* partitioned into two subsets, namely $V_w$, that represents the locations the vehicles may wait, and $V_d$, that represents the locations a service may occur. A vehicle in $V_w$ covers a vertex in $V_d$ if the vehicle can reach the vertex within a specified travel time. Each vertex $u \in V_d$ is associated with a demand that represents a measure of necessity for a vehicle covering the vertex. The objective is for each $k \leq n$ to assign *k* vehicles in vertices of $V_w$, fulfilling a side constraint on the maximum allowed number of relocated vehicles, in order to obtain the maximum expected coverage. The calculation of all $k \leq n$ assignments of vehicles to vertices in $V_w$ is used in order to have all possible relocations known a priori i.e., when a vehicle leaves its position for a service the relocation of the remaining vehicles is already known. The expected coverage is the $\sum_{k=1}^{n} p_k\, c_k$ where $p_k$ denotes the probability exactly *k* vehicles to be available and $c_k$ denotes the coverage obtained from the k vehicles. The problem is formulated as an integer linear programming problem. The approach was tested in real data from Montreal's medical services, with a small number of vehicles $3 \leq n \leq 6$ and the solution was calculated by an integer programming solver. The results indicate that the average response time would not exceed 10 minutes even in the case of only three vehicles. A drawback of the approach of Gendreau et al. is that although they calculate every possible assignment of $k \leq n$ vertices to $V_w$, they do not generate the actual routes to be assigned to vehicles when one of them leaves or becomes available. Furthermore, this approach cannot be used for large values of n, since the solution of the integer programming problem takes exponential time.

Contardo et al. [62] introduce the dynamic public bike-sharing balancing problem. The problem deals with the dynamic relocation of bicycles, i.e. the relocation of bicycles during the day, when the demand for bicycles is not negligible. The aim is to derive vehicle routes for dynamically relocating bicycles in order to minimize the number of occurring shortages, i.e. incidents wherein a customer requires a bicycle from an empty station or attempts to leave a bicycle in a full station. The time is discretized into periods and a space-time network is introduced. A node of the network denotes a station of the bike-sharing system at a specified time period, while an arc represents transition of a station at a specified time period to another station the expected time period. Regarding the number of bicycles loaded on a vehicle (during an arc traversal) as flow, a Mixed Integer Linear Programming (MILP) formulation is proposed. Since, the solution of the previous formulation would be computationally hard, a heuristic approach that produces lower bounds and feasible solutions is also proposed. In the latter two decompositions of the problem are applied. Namely, Dantzig-Wolfe decomposition [68] is applied and the linear relaxation of the formulation is solved, creating a lower bound for an instance. Then a new formulation of the problem is proposed applying Benders decomposition [69]. Taking into account the process applied in the Dantzig-Wolfe decomposition, a feasible solution is obtained. Since no instances for the dynamic relocation of bicycles exist, Contardo et al. [62] created 120 test instances to test their approaches. The instances contained 25, 50 or 100 stations, the time span was set to 2 hours, each time period was set to 5 or 2 minutes and the number of vehicles available for the relocation was considered to be 5. The MILP has been compared against the heuristic approach combining the two decomposition schemes. The former is solved using a commercial solver allowing 30 min execution time. The heuristic approach has been shown to clearly outperform the MILP approach in all test instances apart from the smallest ones. The lower bound produced by the heuristic is higher than the MILP solution, the solution obtained corresponds to lower cost, while the heuristic's execution time does not exceed on average the *6* min, even in the largest instances. On the downside, the problem formulation does not take into account the time spent for loading and unloading bicycles in stations. Since this time is not negligible in relation to the time length of routes and is usually proportional to the number of loaded/unloaded bicycles [61], future work could focus in deriving new formulations of dynamic bicycle relocations incorporating loading/unloading times.

In [25] a bike sharing system consists of a set $S = \{S_1, S_2, ..., S_N\}$ of $N$ stations, where each station $S_i$ has capacity $C_i$ (i.e. it is equipped with $C_i$ bicycle stands). The system employs redistribution to transport bicycles from stations in excess of bicycles to stations that may run out of bicycles. The objective of the control system is for each station $S_i$, to maintain an (appropriate for this station) number of $R_i$ bicycles ensuring that there are always bicycles available for pick up and also $C_i - R_i$ vacant stands available for bicycle drop off. The proposed Petri net model (initially defined only for three stations and generalized to any number of stations in the sequel) consists of three subnets (modules) representing three different functions: (1) the "station control" subnet, representing the control function of the stations to ensure availability of bicycles for pick up and vacant berths for bicycle drop off at every station; (2) the "bicycle flows" subnet representing the bicycle traffic flows between the different stations of the network; and (3) the "redistribution circuit" subnet representing the path to be followed by the redistribution vehicle in order to visit the different stations of the network. The proposed modular and dynamic Petri net model is validated through several simulations made for different interesting system configurations. Labadi et al. argue that Petri nets-based modeling is particularly useful to planners and decision makers in determining how to implement and operate successfully bicycle sharing systems.

Krumke et al. [70] have studied the dynamic relocation problem in car-sharing systems. A customer may pick-up a car from a non-empty station and deliver it to another - not full - station. Similarly to bike-sharing systems, car relocations is necessary to counter the effect of stations with unbalanced vehicle stock. The relocation is assumed to take place using convoys, able to transfer a number of cars between the stations. For each relocation the car system is charged with a cost depending on the number of convoys and the number of cars transferred as well as the distance covered. In the setting of this article a customer reserves in advance, i.e. s/he requests a car rental from a specified station at a certain time to be returned to another specified station at a given time. Furthermore, each request is associated with a profit earned by the system. Based on the previous assumptions two variants of the relocation problem are tackled. In the first, all requests must be serviced and the goal is to minimize the cost of the relocation operations to meet all customers' demand. In the second, the goal is to decide which requests to service as well as to schedule the relocation tours of convoys to maximize the system's profit. An integer linear program is introduced for each of these variants. The solution approach incorporates a time-expanded network. The network consists of nodes representing stations at specified times. Network arcs represent feasible transitions of convoys with cars between station-time pairs. Arcs also model requests information i.e., a transition of a rent car from a station to another. Two kinds of flows are introduced. A flow representing the number of cars transferred between stations and a flow representing the number of convoys moved between stations. These two flows are related to each other, by introducing a constraint on the maximum number of cars transferred with a specified number of convoys, based on the capacity of the convoys. Based on these flow considerations, two integer linear programming formulations are proposed for the investigated problems. Notably, solving the proposed integer programs is highly inefficient, yet, no efficient algorithm is proposed by the authors to meet the requirements of the dynamic relocation scenario. Furthermore, the use of convoys for transferring cars may not be feasible in many urban settings and, hence, employed drivers may be restricted to transfer at most one car at a time.

Lee et al. [71] have studied the static relocation problem in EV-sharing systems. The authors assume concurrent relocation of EVs (i.e. employed drivers are assumed to be at the stations where cars to be relocated reside at the time that relocation starts) without taking into account their residual energy. The relocation policy aims to restore a certain availability of vehicles at each station. They consider both even relocation schemes (stations end up having equal number of cars) and utilization-based relocation schemes (EVs are assigned to stations according to the demand ratio of each station, known a priori). Depending on the chosen relocation scheme a relocation vector is determined, namely, the desired number of vehicles in each station after the completion of the relocation process. A heuristic algorithm is executed thereupon, deciding the destination of each vehicle through matching EVs from overflow stations to underflow stations so as to minimize the relocation cost, i.e. the overall moving distance. For an EV, the preference to underflow stations depends on the distance to be travelled. Each EV to be reallocated has an index to its preference list of underflow stations, with the index initialized to the first (i.e. nearest one). In addition, each underflow station is aware of its required number of EVs and maintains an allocation list (i.e. list of EVs residing elsewhere and currently assigned to it). EV-station matching begins from the first EV (among the ones scheduled to be relocated). The EV examines the option of relocating to the station marked by its local index within its preferences list. If the station currently holds within its allocation list less EVs than the required number e, the EV is added to the allocation list. Otherwise, if the allocation list is full, the EV which is farthest away from the station is removed. The removed EV then examines the next station, shifting ahead the index in its preference list. Having completed this iterative process, the allocation lists of underflow stations should be finalized. Such allocation, represented by (EV, station) pairs, should have the minimum relocation distance.

In a follow-up work, Lee and Park [72] designed a team-based relocation scheme for EV-sharing systems and proposed a genetic algorithm-based solution to obtain a reasonable quality relocation plan within a limited time bound. Each relocation plan, namely, the set of relocation pairs of EV from overflow to underflow stations, is represented by an integer-valued vector to run the genetic operators such as crossover, selection, reproduction, and mutation. Drivers performing vehicle relocations are assumed to move in teams of *m* members, wherein one of them drives a car following a route through a series of overflow stations. Upon arriving at a station, the rest of the team members drive *m-1* EVs to the (same) planned underflow station, while the driver of the relocation vehicle follows them. This process is repeated until the relocation procedure completes. The experimental results have demonstrated that each addition of a service staff may significantly decrease the relocation distance.

The previous algorithmic approaches are summarized in Table 4 where a classification is also given according to whether they are static or dynamic, bicycle or car repositioning techniques.

|  | Static bike repositioning | Static car repositioning | Dynamic bike repositioning | Dynamic car repositioning |
|---|---|---|---|---|
| One commodity capacitated pick up and delivery formulation + Heuristic approach | Chemla et al. [63] |  |  |  |
| MILP formulation + Heuristic approaches | Raviv et al. [61] | Kek et al. [30] | Contardo et al. [62] |  |
| Maximal Expected Coverage Relocation Problem / ILP approach |  |  |  | Gendreau et al. [67] |
| Petri net modeling approach |  |  | Labadi et al. [25] |  |
| ILP approaches |  |  |  | Briest and Raupach [66] Krumke et al. [70] |
| Heuristic approaches |  | Lee et al. [71] Lee and Park [72] |  |  |

**Table 4**. Algorithmic approaches on the operational repositioning of bicycles/cars

## 4. Algorithms for Ride Sharing

Ride sharing is promoted as a way to better exploit unused car capacity, thus lowering fuel usage and transport costs. In the context of a vehicle sharing system, ride sharing can be used to maximize the profit of the system by further optimizing cars usage and minimizing the number of unsatisfied customer requests in the case that there are no available cars in certain pickup stations and/or parking slots in drop-off stations. In this Section we summarize algorithmic approaches that deal with challenges arising in the domain of ride sharing, in particular the proper assignment of driver's offers and requests in ride sharing applications. All techniques aim at fast running times to allow real-time applications.

In [54] Geisberger et al. provide practical algorithms to compute detours in the context of ride sharing. They consider the scenario where queries of users wishing to get from an origin *s* to a destination *t* should be matched to offers from riders going from *s'* to *t'*. Two types of possible matches are distinguished. In case of a perfect fit, the sources *s, s'* and destinations *t, t'* of driver and rider, respectively, are identical. In a reasonable fit, small detours and additional stops are allowed. The goal is to find the offer for which the detour is minimized. Formally, the goal is to minimize *d(s', s) + d(s, t) + d(t, t') − d(s', t')*. The authors present an algorithmic approach to find reasonable fits for a set of offers and a single incoming request by using Dijkstra's algorithm [55] to compute the detour for each offer, and return the offer with minimum detour.

Using a well-known speed-up technique called Contraction Hierarchies [56], Geisberger et al. are able to achieve query times that are faster than the straightforward approach described above. This alternative approach exploits the structure of search spaces in Contraction Hierarchies. The search space consists of two independent parts, namely the forward and the backward search space. More specifically, assuming that there are *k* offers, for every incoming *s-t* request, *k* queries from *t* to $t_i'$ need to be run (one for each offer $t_i'$). However, all these queries have exactly the same forward search space, so the forward search space only needs to be computed once. In addition to that, the results of backward searches can be precomputed for each offer $t_i'$. Each vertex in the backward search space of $t_i'$ gets a bucket assigned to store the corresponding distances. Experiments show that using these techniques allows to answer incoming queries several orders of magnitude faster than the straightforward Dijkstra-based approach.

In [57] Abraham et al. present a fast algorithm, HLDB, to compute shortest path distances using preprocessed data based on hub labels [58]. Hub labels are sets of "important" vertices of a graph *G = (V, E)*. Each vertex *v* ∈ *V* has a forward label $L_f(v)$ and a backward label $L_b(v)$. For each hub vertex *h* ∈ $L_f(v)$, they precompute and store the distance *d(v, h)* from *v* to *h*. An *s-t* distance query then checks for a hub *h* ∈ $L_f(s)$ ∩ $L_b(t)$ that minimizes the distance *d(s, h) + d(h, t)*. To preserve correctness, the labels must fulfill the *cover-property*, that is, for any pair *s, t* ∈ *V*, $L_f(s)$ ∩ $L_b(t)$ must contain a vertex on a shortest path from *s* to *t*. Precomputing labels that fulfill these properties can be accomplished using a technique based on Contraction Hierarchies [56]. Several heuristics are added to improve the performance of the algorithm.

One special property of the technique presented in [57] is that it works entirely with a database, using SQL queries. Although their basic case considers peer-to-peer shortest path queries, they consider several extended scenarios, such as POI-queries and ride sharing. The scenario mentioned above is examined, where queries and offers are to be matched. Again, the goal is to find an offer for which the detour is minimized, i.e., that minimizes *d(s', s) + d(s, t) + d(t, t') − d(s', t')*. Using the HLDB approach, the authors show how to solve this problem efficiently with simple database operations. To answer queries, a table *offers* is created containing the four columns *id, source, target,* and *distance*. A second table *offers_labels* contains the four columns *id, hub_forward, hub_backward,* and *distance*. For every offer *(s', t')*, each combination *(h, h')* stores an offer *ID* in *id*, *h'* in *hub_forward*, *h* in *hub_backward* and the distance *d(s', h) + d(h', t') − d(s', t')* in *distance*. Computing the minimum distance can now be done with loops over all possible combinations.

In order to allow for more flexible scenarios, Drews and Luxen [59] introduce multi-hop scenarios, where users can even transfer between cars of different drivers. Such transfers occur at designated stations $S \subseteq V$ (e.g., parking lots). Their scenario extends the previous approaches by adding time-dependency. Offers and requests are given as triples *(s, t, τ)*, where *s* and *t* are vertices, and *τ* is a departure time. Given all stations $s_i$ at which transfers occur and possible waiting times $\omega_m(s_i)$ for a match *m*, the total duration of a journey equals

$$d(m) = \sum_{i=0}^{t} (\omega_m(s_i) + d(s_i, s_{i+1}))$$

One way to model the resulting time-dependent scenario is to use time-dependent graphs [60]. The authors extend this model by introducing the so-called "Slotted Time-Expanded Graphs". Here, the continuous time divided into equal-sized time slots, and departures are assumed only to happen at the end of such slots. This results in a directed acyclic graph, where finding the best fit for a certain request is done by running Dijkstra's algorithm. An *A\**-variant is used to achieve speedup by about two orders of magnitude. The experimental study also evaluates the quality of their solutions and shows that request and offers are well matched by the proposed techniques.

## 5. Research Challenges and Future Prospects

The viability of bicycle/car sharing systems largely depends on their effective strategic design, management and operation. Along this line, three separate planning aspects are identified: (i) strategic (long-term) network design comprising decisions about the location and the number of stations as well as the vehicle stock at each station; (ii) tactical (mid-term) incentives for customer-based distribution of bicycles/cars, i.e. incentives given to users so as to leave their vehicle to a station different to that originally intended (this may be regulated through pricing schemes adaptable to the system state); (iii) operational (short-term) operator-based repositioning of bicycles/cars based on the current state of the stations as well as aggregate statistics of the stations' usage patterns.

However, each of the abovementioned planning aspects raises considerable research challenges:

- Strategic design should balance among the system's intended quality of service and investor cost. On one hand, the investor's cost includes the facility costs of bike docks or dedicated parking stations, acquisition and maintenance costs for vehicles, vehicle depreciation costs and routine relocation costs. On the other hand, the level of service provided by the system mainly depends on vehicles' availability to satisfy user demand distribution in space and time.
- Incentives should be carefully designed so as to align the travel behavior of the users with the system's pursuit, which dynamically adapts to real time demand patterns. In particular, the incentivisation scheme should aim at increasing the number of served customers, offer meaningful and attractive alternatives to incentivized customers and minimize the revenue losses for the operator.
- The operational repositioning of vehicles differs considerable among bike and car-sharing systems. In the former, bikes are carried in bunches by dedicated vehicles whereas in the latter cars are relocated by groups of drivers. Both represent tough optimization problems, wherein the objective is to minimize relocation cost while satisfying user demand. Furthermore, the effectiveness of vehicle relocation schemes should ideally consider future demand so as to proactively ensure the availability of vehicles areas in which demand is expected to rise.

In the sequel of this Section open research issues relevant to the abovementioned issues are identified.

**Strategic design of bicycle sharing systems.** The design of bicycle sharing systems is a particularly tough exercise as it involves several design decisions: the location and capacities of bike stations; the vehicles fleet size; the creation of bicycle lanes connecting bike stations. These design decisions should ensure sufficient coverage (i.e. satisfy user demand with appropriate quality of service), while minimizing investment and maintenance costs. Despite the growing body of relevant literature, algorithmic approaches tackling this issue are very few. Given the problem's complexity and the metropolitan scale of realistic instances, heuristics represent a suitable method for efficiently deriving near-optimal solutions. Existing algorithmic approaches on hub location problems, maximal covering models and joint location inventory problems are expected to provide a suitable starting point for optimizing the design of bicycle sharing systems. Along the same line, the mathematical modeling of bicycle sharing systems should be refined so as to capture several system variables and constraints overlooked by existing models:

- Travel demands may largely vary over a day (e.g. residential areas typically act as trip generators early in the morning and attractors late in the afternoon). It would therefore be helpful to develop a formal model incorporating demand variation and to evaluate the influence of demand variation on the system design and routing choices.
- Decisions on the establishment of new bicycle lanes between bike stations should take into consideration the existing street network structure (unlike existing models which simplistically consider direct links between stations [26]). Clearly, existing bicycle lanes infrastructure should be exploited in order to reduce facilities cost. Moreover, new lanes should be setup taking into account the attractiveness of alternative options with respect to distance, bicycle friendliness (e.g. road segments with high motor traffic are less friendly to bikers than pedestrian zones), flatness, etc. Notably, OpenTripPlanner (OTP[7]), the leading open source platform for multimodal trip itinerary planning, already supports the provision of such information[8].
- The reallocation of bicycle stock is commonly practiced by shared bicycle system operators to enable balanced distributions among stations and allow coverage of anticipated demand. Given that bicycle reallocation largely contributes to maintenance cost, model formulations should consider this cost factor so as to influence the overall system design.

**Mobility on Demand combined with dynamic incentives.** It seems that incentives-based mechanisms represent the most promising way to deal with the major problem of car sharing systems, that is, the fleet redistribution issue or asymmetric demand/offer. In effect, incentives motivate fleet redistribution and tackle the demand/offer asymmetry problem. This mechanism is based on real time bi-univocal information between the user and the system, allowing to modify, not only the drop off and pick up station, but also other trip parameters such as the routing options for moving from a location *A* to a location *B* the time for picking up or dropping off the car, suggest trip sharing with another user going along the same route, etc. Thus, it might occur that moving from a location *A* to a location *B* has different prices depending on the incentives or penalties offered and accepted by the user. Incentives must be managed in real time and the system should be adaptive and possess some kind of intelligence to infer/plan each user behaviors/tendencies, so that, it is able to offer a particular user the "right" incentive (i.e. attractive enough for the user to modify his/her initial plan but adjusted enough to maximize the benefit of the fleet manager). Incentives may be offered in two forms: in kind or in price. "In kind" incentives refer to discount vouchers or special offers for services -directly or indirectly- relevant to mobility. For example, it could be a 15% discount on a restaurant, or free laundry service or allowance to top price range vehicles in the system. Of course, there should be previous agreement among cooperating establishments (offering these "in kind" incentives) and the fleet management authority. Price incentives refer to discounts on actual or future trip fares and exclusively involve fleet management services. Finally, taking the incentives scheme to the extreme, there could be a way to make it explicit to the users. When the asymmetric demand problem deteriorates, the fleet manager could "offer" to users - deliberately subscribed for this purpose - an attractive incentive to drive a vehicle from *A* to *B*. The user answering positively would earn future discounts or even monetary rewards for driving the car from *A* (place with low demand) to *B* (place with high demand). This option could be seen as a contractor-based redistribution system. However, the use of this incentive tactic should be implemented in severe asymmetry situations because of the high "redistribution" trip costs incurred by the fleet manager.

---

[7] http://opentripplanner.com/
[8] OpenTripPlanner relies on General Transit Feed Specification (GTFS) (https://developers.google.com/transit/gtfs/) data to describe public transportation schedules and routes. It can use OpenStreetMap (http://www.openstreetmap.org/) or commercial data sources for data on sidewalks, bicycling infrastructure and streets. It allows users to plan a trip that can combine multiple modes of transportation, such as cycling or walking to reach public transportation, while it can also incorporate several popular bike-sharing systems (http://wiki.openstreetmap.org/wiki/OpenTripPlanner).

**Vehicle relocation and effective reward schemes.** Car relocation is deemed as a necessary instrument to restore the desirable allocation of vehicles among stations in car-sharing systems. Having to adapt to user demand dynamics, car relocation activities are typically needed several times on the course of a day; hence, relocation decisions are bound to time constraints. Given the complexity of the problem, heuristics represent a reasonable algorithmic tool to meet the strict time requirements. However, the algorithmic state of the art in dynamic vehicle relocation in car-sharing systems leaves a lot to be desired. For instance, the results obtained by the greedy approach of Lee et al. [71] could be significantly improved by approaching car relocation as a *k*-server problem (regarding the employed drivers of the car sharing operator as servers that handle relocation requests). Moreover, the problem of optimally assigning employed drivers to cars to be relocated and transferring the drivers to the stations where those cars reside has not been studied, although being an essential part of the relocation process. The provision of incentives to customers has also been recognized as a cost-effective means of tackling the problem of unbalanced car distribution among stations in car sharing systems. The benefit of incentive provision models has been evidenced by several simulation studies (see Subsection 3.2). In real-world systems, though, users indicating willingness to take advantage of a reward scheme would expect meaningful alternatives. For instance, a customer would consider delivering a car to a station further than that originally planned, under the condition that s/he could transfer to a transit service and reach his actual destination location with reasonably small delay. Furthermore, such meaningful recommendations should maximize the utility for the system (e.g. incentivize the customer to undertake the most urgent, among pending, relocation) and should be derived in real time. Last, rewards (i.e. rental discount) need to be adjusted so as to compensate the user enough for delaying his/her arrival time (or even having to pay for a transit service ticket), while minimizing revenue cost for the operator. To the best of our knowledge, no algorithmic methods have been proposed so far deriving concrete alternatives so as to effectively incentivize customers. Hence, this represents a particularly promising research topic.

**Proactive vehicle relocation based on predicted demand.** Contemporary vehicle sharing systems take a reactive approach to handling user demand, wherein vehicles are relocated from station with surplus to those with shortage of vehicle stock, as soon as uneven vehicle distribution is detected. Given the highly dynamic nature of user demand, such relocations are likely to prove ineffective, e.g. relocate vehicles to stations with relatively low stock and yet to remain unused. The use of historical data and demand prediction models may, however, give effect to more effective relocation strategies. For instance, a vehicle depot located nearby office premises with a few parked cars may be reasonable to supply before the end of the business hours. This relocation may be undertaken either by operator employees (relocators) or incentivized customers. In the special case of EV-sharing systems, demand prediction may be used to identify which vehicles (among those parked at a specific depot) should be relocated; for instance, vehicles with high battery level may be more appropriate to relocate to a station at a time that relatively long rides are expected to be requested. Furthermore, the limited range and the long charging of EVs give reasons to innovative incentivized schemes. For instance, in the event of a request issued at 20pm for a 25km ride towards a suburb where high user demand is not expected before 7am, the customer could be incentivized to use a vehicle with battery status providing 35km autonomy, which requires 8 hours to be fully charged.

## 6. Conclusions

Vehicle sharing represents an emerging transportation scheme which may comprise an important link in the green urban mobility chain. One-way vehicle sharing systems employ a flexible rental model (customers are allowed to pick-up a vehicle at any station and return it to any other station) which best suits typical urban journey requirements. However, the so-called demand-offer asymmetric problem (i.e. the unbalanced offer and demand of vehicles) typically experienced in one-way sharing systems severely affects their economic viability as it implies that considerable human (and financial) resources should be engaged in relocating vehicles to satisfy customer demand.

The design and management measures aiming at alleviating imbalances in the availability of bicycles/cars, are distinguished into three separate planning horizons [23]: (i) Strategic network design comprising decisions about the location and the number of stations as well as the fleet size at each station, (ii) tactical incentives for customer-based distribution of bicycles/cars and (iii) operational (operator-based) repositioning of bicycles/cars based on the current state of the stations as well as aggregate statistics of the stations' usage patterns. This chapter presents an extensive literature survey on models and algorithmic techniques for the design, operation and management of vehicle sharing systems. Different approaches applied either to bike or to car sharing systems, are described and classified according to the involved solution method. Also, open research problems relevant to the abovementioned issues are identified highlighting important research issues that need to be addressed in the future.

**Acknowledgment**

This work has been supported by the EU FP7 Collaborative Project MOVESMART (Grant Agreement No 609026).